\documentclass[aps,showpacs,onecolumn,floats,prd,superscriptaddress,nofootinbib]{revtex4} 
\usepackage{graphicx,amsmath,amssymb,amstext}
\usepackage{amssymb,amsbsy,amsfonts,amsthm,color}

\usepackage{epsfig}
\usepackage{graphicx}
\usepackage{subfigure}

\graphicspath{{Figures/}}

\begin{document}

\title{Pulsar Acceleration Shifts from nearby Supernova Explosion}

\author{Darsh Kodwani}
\email{dkodwani@physics.utoronto.ca}
\affiliation{Canadian Institute of Theoretical Astrophysics, 60 St George St, Toronto, ON M5S 3H8, Canada.}
\affiliation{University of Toronto, Department of Physics, 60 St George St, Toronto, ON M5S 3H8, Canada.}

\author{Ue-Li Pen}
\email{pen@cita.utoronto.ca}
\affiliation{Canadian Institute of Theoretical Astrophysics, 60 St George St, Toronto, ON M5S 3H8, Canada.}
\affiliation{Canadian Institute for Advanced Research, CIFAR program in Gravitation and Cosmology.}
\affiliation{Dunlap Institute for Astronomy \& Astrophysics, University of Toronto, AB 120-50 St. George Street, Toronto, ON M5S 3H4, Canada.}
\affiliation{Perimeter Institute of Theoretical Physics, 31 Caroline Street North, Waterloo, ON N2L 2Y5, Canada.}

\author{I-Sheng Yang}
\email{isheng.yang@gmail.com}
\affiliation{Canadian Institute of Theoretical Astrophysics, 60 St George St, Toronto, ON M5S 3H8, Canada.}
\affiliation{Perimeter Institute of Theoretical Physics, 31 Caroline Street North, Waterloo, ON N2L 2Y5, Canada.}

\begin{abstract}
We show that when a supernova explodes, a nearby pulsar signal goes through a very specific change. The observed period first changes smoothly, then is followed by a sudden change in the time derivative. A stable millisecond pulsar can allow us to measure such an effect. This may improve our measurement of the total energy released in neutrinos and also the orientation of the supernova-pulsar system. 
\end{abstract}

\maketitle

\section{Introduction and Summary}

When a core collapse supernova (SN) explodes, it releases approximately a fraction of the solar-masses worth of energy in a shell of relativistic neutrinos. Although neutrinos barely interact with other matter directly, relocating such amount of mass inevitably affects the background geometry. If an accurate timing apparatus is nearby, such as a millisecond pulsar, then one can hope to observe an effect. The purpose of this paper is to calculate such an effect and provide a better idea of how useful such an observation can be.

Since this neutrino shell is very thin---passing through either the Earth or the pulsar in a time much shorter than the typical duration for timing observations---we can treat it as a co-dimension-1 delta-function and describe the spacetime by Israel Junction Conditions (IJC) \cite{Isr66}. Furthermore, we will treat both the neutrinos and the pulsar signal as propagating at the speed of light. In reality that is not exactly true, with photons moving even slower than neutrinos. Nevertheless, the time delay is again much shorter compared to the duration of timing observations, thus it does not invalidate our result. The analytical model then becomes quite simple, since it involves just tracking geodesics across a thin shell \cite{BouFre07,JohYan10,OluPie13}.

In Sec.\ref{sec-JC}, we present the general setup of geometry with junctions. In Sec.\ref{sec-1+1} we study the scenario in which the SN and the pulsar are exactly aligned, as shown in Fig.\ref{fig:1}. On one hand, this is the simplest scenario as it reduces to a (1+1) dimensional problem. On the other hand, this might have the strongest observable effect. After all, the signals that arrive right before and right after the neutrino shell travelled in two dramatically different geometries. We show that the observed pulsar period does not jump across the neutrino shell. In hindsight, these should not be too surprising, as they follow from the physical continuation of spacetime. Although the metrics on either side of the neutrino shell look very different, the physical distance between two points, and the relative velocity between two points, cannot change discontinuously. 
Thus the leading observable effect is in the derivative of the pulsar period, due to the change in its gravitational acceleration. 
\begin{equation}
\delta(\dot{P}/P) \sim -\delta M / r_p^2~,
\label{eq-main}
\end{equation}
where $P$ is the observed pulse period, $G$ and $c$ are set to 1, $\delta M$ is the total neutrino shell mass thus also its effective Schwarzschild radius, and $r_p$ is the distance between the SN and the pulsar. 

In Sec.\ref{sec-3d}, we study the general case in which the pulsar and SN are not aligned. The above acceleration change happens when the shell hits the pulsar. However, before seeing that, we will first see the SN explosion, and thereafter the pulses must cross the neutrino shell before reaching us. Such shell-crossing also induces observable changes in the pulsar period at similar order of magnitudes \cite{OluPie13}. Combining these two effects, we derive the a unique signature in pulsar observation accompanying the SN explosion.

In Sec.\ref{sec-obs}, we show that the required timing accuracy to measure such a signature is the time it takes light to cross the effective Schwarzschild radius of the neutrino shell. That is about $10^{-6}$ seconds, which is achievable by millisecond pulsars. The particular feature in the time-shifts can determine the total energy released in neutrinos and the orientation of the SN-pulsar system.

\section{Geometry}
\label{sec-JC}

We will calculate the observable effect by treating the spacetime around the SN as roughly a Schwarzschild geometry, while the pulsar is like a test particle in this background.
\begin{equation}
	ds_i^2 = -\left( 1 -\frac{2M_i}{r}\right) dt_i^2 + \left( 1 -\frac{2M_i}{r} \right)^{-1} dr^2 + r^2 d \Omega_2^2~.	\label{2.1}
\end{equation}
Here the $i$ index on quantities stands for the ``initial" spacetime, thus $M_i$ is the total mass of the SN (before it loses the neutrino shell). Similarly, replacing $i$ by $f$ means the final spacetime where $\delta M = (M_i-M_f)$ is the total energy carried away by the neutrino shell. In principle every coordinate should have a subscript $i$ or $f$. However, since we assume spherical symmetry, we can identify their angular coordinates and declare the radial coordinate as the area radius of the two-sphere. Thus, the only coordinate difference appears in $t_i$ and $t_f$. 

The initial and final geometries are connected through a junction, which is the neutrino flow that we treat as a thin, null shell. The radial 4-vector of this shell is thus a null vector, which can be represented in either coordinate.
\begin{eqnarray}
j_\mu &\equiv& \left( \left(1-\frac{2M_i}{r}\right)^{-1} ,  1 , 0 , 0 \right)_{\rm in \ metric \ i}~.
\end{eqnarray}
Throughout this paper, we will use ``$\equiv$'' to represent a physical quantity in a coordinate system. In the ``final" metric, one replaces the ``$i$'' in the above expression by ``$f$''. That leads to a mathematically different component form, but it represents the same physical 4-vectors.

We will treat our signal as photons and talk about their change of frequencies. In reality, one just relates this to the pulsar period by $w=(2\pi/P)$. A photon can be described by a null-4-vector with a fixed normalization as shown in section 25.6 of \cite{MTW}. In a fixed Schwarzschild geometry, the general form is
\begin{equation}
	k^\mu \equiv w_{\infty}\left( \frac{1}{(1- \frac{2M}{r})}, \sqrt{ 1 - \frac{b^2}{r^2} \left( 1 - \frac{2M}{r} \right)}, \frac{b}{r^2}, 0 \right)~. \label{eq-photon}	
\end{equation}
The overall normalization $w_{\infty}$ is defined as the frequency observed by an asymptotic observer at rest, and the fixed angular momentum is parametrized by the impact parameter $b$. For convenience, we can always assume that the motion is confined to the equator of the two sphere.

\section{The Aligned Case}
\label{sec-1+1}

\begin{figure}[tb!]
\begin{center}
\includegraphics[scale = 0.3]{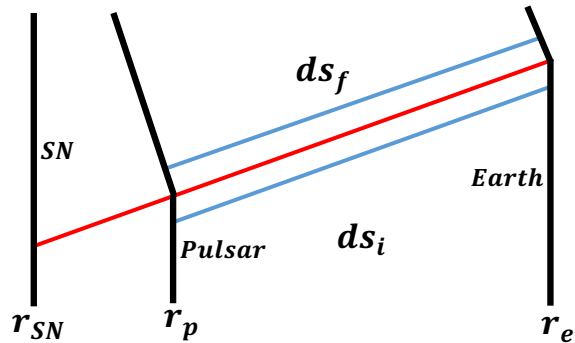}
\caption{Spacetime diagram showing the position of the pulsar, $r_p$, earth, $r_e$, and the SN, $r_s$. The red line represents the neutrino null vector and the blue lines represent the photon trajectories.}
\label{fig:1}
\end{center}
\end{figure}

First we will consider the case that $b=0$ in Eq.~(\ref{eq-photon}), namely the SN, the pulsar, and the Earth sit on one straight line, and the pulsar is in the middle. This reduces to a (1+1)-dimensional problem as depicted in Fig.\ref{fig:1}. In this scenario, the earth will receive two photons from the SN, one right after the other, but they have traveled through two distinctly different geometries. One might suspect a sudden and dramatic change in the signal, but it is easy to show otherwise.

Let $p_\mu$ and $e_\mu$ be the timelike 4-vectors of the pulsar and the earth while emitting/receiving these pair of photons. Continuity across the junction implies the following constraint on how they are represented in either geometry.
\begin{eqnarray}
p_\mu j^\mu(r_p)_{\rm \ in \ metric \ i} &=& 
p_\mu j^\mu(r_p)_{\rm \ in \ metric \ f}~,
\label{eq-JuncPuls} \\ 
e_\mu j^\mu(r_e)_{\rm \ in \ metric \ i} &=& 
e_\mu j^\mu(r_e)_{\rm \ in \ metric \ f}~.
\end{eqnarray}
The two photons are defined by the physical fact that in the rest frame of the pulsar, they have the same frequency.
\begin{equation}
w_p = -p_\mu k^\mu(w^i_\infty, r_p)_{\rm i} = 
-p_\mu k^\mu(w^f_\infty, r_p)_{\rm f}~.
\label{eq-PulsEmit}
\end{equation}
It is then straightforward to combine these 3 equations to show that the frequencies received on earth are identical.
\begin{equation}
w_e = -e_\mu k^\mu(w^i_\infty, r_p)_{\rm i} = 
-e_\mu k^\mu(w^f_\infty, r_p)_{\rm f}~.
\end{equation}

\subsection{Leading order expansion and physical intuition}

Our conclusion above is exactly correct in full nonlinear general relativity, although it might be surprising to some readers. Here we will expand the two steps separately to leading order for better physical intuitions. First of all, it is quite realistic to set $r_e \gg r_p$ and treat it as being infinite, so the only physical change happens at the pulsar.

Instead of using Eq.~(\ref{eq-JuncPuls}), if we had assumed that the pulsar is at rest in both coordinates before and after the neutrino shell crosses, then we would have derived a frequency ratio of $[1-\delta M/r_p]$ between these two photons\footnote{In fact, any non-relativistic initial velocity leads to this result.}. This would have reflected the fact that they climbed out of two different gravitational potentials, but it cannot be the full story. If the pulsar was at rest before the neutrino shell hits it, Eq.~(\ref{eq-JuncPuls}) demands that it picks up a velocity in the new coordinate
 \begin{eqnarray}
p_\mu &\equiv& \left( \left( 1 - \frac{2M_i}{r} \right)^{-\frac{1}{2}}, 0, 0,0\right)_{\rm i} 
\label{eq-PulsarRest} \\ \nonumber  &\equiv&	
\left( \left( 1 - \frac{2M_f}{r} \right)^{-\frac{1}{2}}, -\frac{\delta M}{r_p},0,0 \right)_{\rm f}~.
\end{eqnarray}
The new velocity, $v=\delta M/r_p$, is falling toward the SN remnant. This then creates a Doppler shift that exactly cancels the difference in gravitational potential, thus netting no frequency change.\footnote{This change in coordinate velocity is not due to any direct interaction with the neutrino shell, but simply an artifact of coordinate change across the junction.}
\begin{equation}
(w_e)_{\rm i} \equiv w_p \left(1-\frac{M_i}{r_p}\right) = 
w_p (1-v)\left(1-\frac{M_f}{r_p}\right) \equiv (w_e)_{\rm f}~.
\label{eq-leading}
\end{equation}

\subsection{Observable change in acceleration}
\label{sec-acceleration}

Let us continue the approximation in the previous section and take a time derivative of Eq.~(\ref{eq-leading}). Realistically, $v \ll1$ and $r_p\gg M$, then the dominant change in time comes from the velocity.
\begin{equation}
\frac{\dot{w_e}}{w_e} = -\dot{v} = -\frac{M}{r_p^2}~.
\label{eq-AccChange}
\end{equation}
When the pulsar crosses the neutrino shell, the enclosed mass changes from $M_i$ to $M_f$, and $\dot{w_e}$ indeed changes suddenly. It is not a suddenly observable change, since it takes some time to build up the change in $w_e$ and then in the phase before it is observable. Nevertheless, this is the leading order observable for us.

\section{General Cases}
\label{sec-3d}

\begin{figure}[t]
\begin{center}
\includegraphics[scale = 0.3]{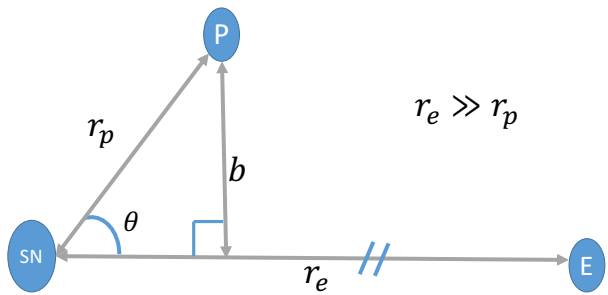}
\vspace{-8 mm}\caption{General geometry of the three objects. Where $b$ is the impact parameter of the photons from pulsar to earth. } 
\label{fig:3}
\end{center}
\end{figure}

In general, the three bodies in this problem will not exactly align. Let $\theta$ be the angle between $r_p$ and $r_e$ as shown in Fig.\ref{fig:3}, we will calculate the general change in acceleration under the same assumption that has a convenient 1st order expansion: $r_e\gg r_p\gg M$ and a non-relativistic initial velocity.

\begin{figure}[tb]
\begin{center}
\includegraphics[scale = 0.3]{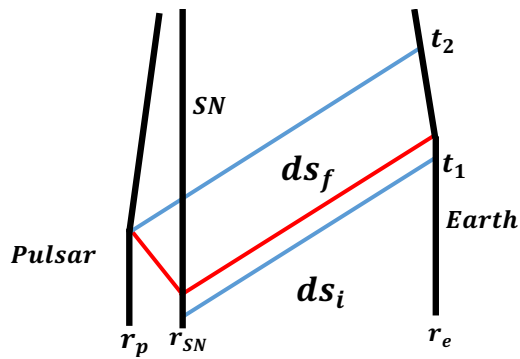}
\caption{Spacetime diagram showing the pulsar at $\theta \approx \pi$ with the same color-code as Fig.\ref{fig:1}. The two photons are right before the first change and right after the second change of $\dot{w_e}$.}
\label{fig-opposite}
\end{center}
\end{figure}

Recall that in the aligned case, there was one special ``time'' (a particular photon in the train of signals from the pulsar), before which both the earth and the pulsar are outside the neutrino shell, and after which they are both inside it. In the general case, there are two such special ``time''s, and this is easiest to understand from the special case in which $\theta \approx \pi$, as shown in Fig.\ref{fig-opposite}. $t_1$ is when the neutrino shell passes through the earth, which is also when we observe the SN explosion. After that, every pulse must cross the neutrino shell somewhere in its path before reaching us. Such crossing lasts until $t_2$, when the neutrino shell hits the pulsar itself. From the earth's perspective, that is when we see the first photon from the pulsar after it is hit by the neutrino shell.

Before $t_1$, both the pulsar and the earth are in metric $i$. After $t_2$, both are in metric $f$. Thus we simply have
\begin{eqnarray}
\frac{\dot{w}_e}{w_e} &=& -\frac{M_i}{r_p^2}\cos\theta~, \ \ \ {\rm for} \ \ t<t_1~,
\label{eq-before} \\
&=& -\frac{M_f}{r_p^2}\cos\theta~, \ \ \ {\rm for} \ \ t>t_2 = t_1 + r_p(1-\cos\theta)~,
\end{eqnarray}
which comes from the projection of gravitational acceleration in our direction. Between $t_1$ and $t_2$, a pulse originates in metric $i$, but must cross the neutrino shell (being caught-up by it) and enter metric $f$ before reaching the earth. This shell-crossing leads to a continuous change in the observed frequency. The change in photon frequency when the pulsar remains stationary was calculated in \cite{OluPie13}. Here we include the fact that before $t_2$, the pulsar is still feeling the same acceleration which lead to Eq.~(\ref{eq-before}). Without loss of generality, we set $t_1=0$, and the full change in frequency through out the process is then given by
\begin{eqnarray}
w_e(t) &=& w_0 \left[ 1 -  t\frac{M_i}{r_p^2}\cos\theta \right]~, 
\ \ \ {\rm for} \ \ t < 0~, \label{eq-beforeW}  \\
&=& w_0
\left[ 1 -  t\frac{M_i}{r_p^2}\cos\theta + \frac{4\delta M t^3}{(t^2 + r_p^2\sin^2\theta)^2} \right]~, 
\ \ \ {\rm for} \ \ 0 < t <  r_p(1 - \cos\theta)~, \\
&=& w_0
\left[ 1-\frac{M_i}{r_p}\cos\theta(1-\cos\theta) + \frac{\delta M}{r_p}(1-\cos\theta) -  \bigg(t-r_p(1-\cos\theta)\bigg) \frac{M_f}{r_p^2}\cos\theta \right] \nonumber \\
& & {\rm for} \ \ t > r_p(1-\cos\theta)~.
\end{eqnarray}
Here $w_0 = w_e(0)$ is the frequency observed on earth at the moment when we see the SN explosion. 

\section{Observational Practicality}
\label{sec-obs}

\begin{figure}[t]
\includegraphics[width=0.3\textwidth]{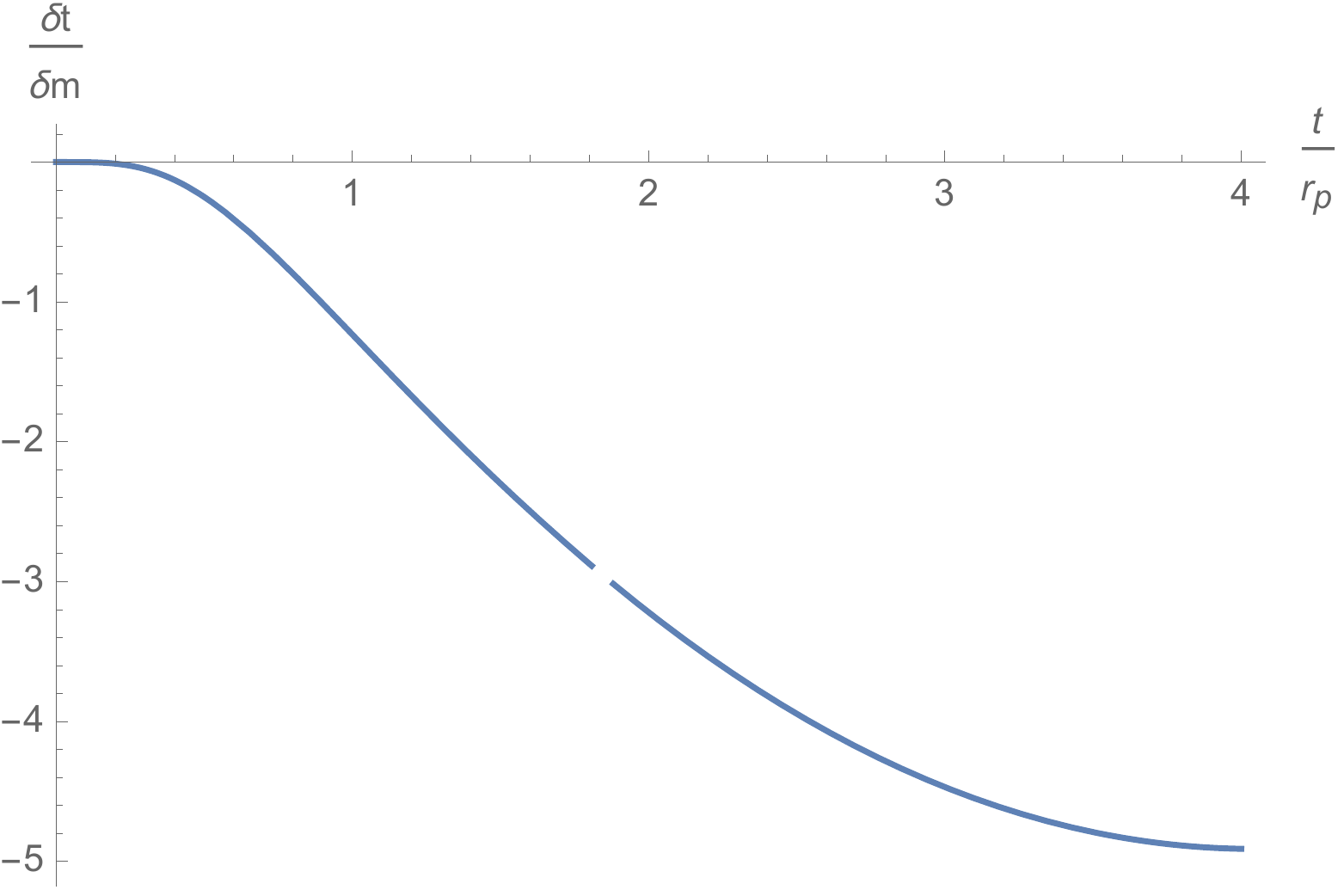}
\includegraphics[width=0.3\textwidth]{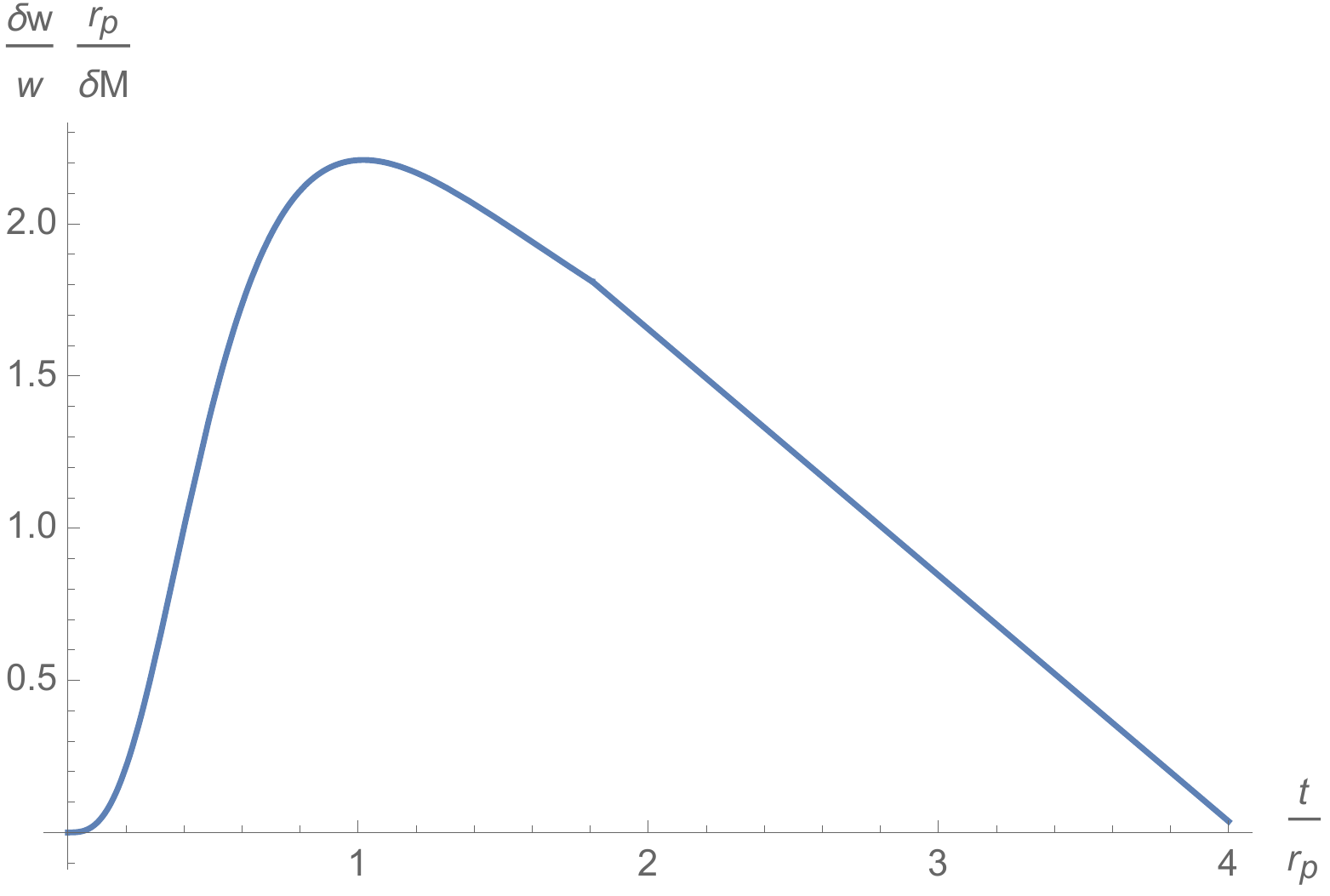}
\includegraphics[width=0.3\textwidth]{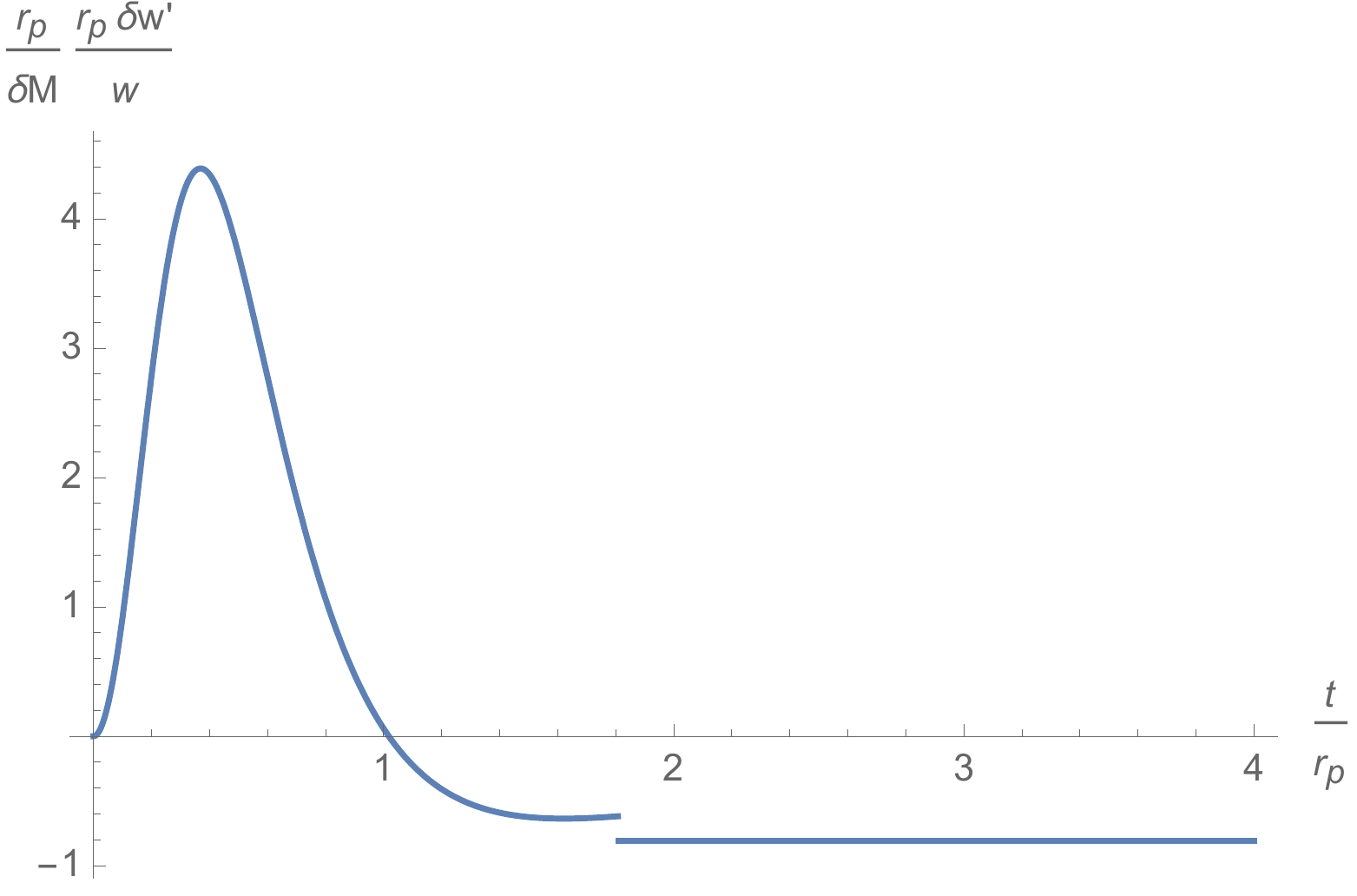} \\
\includegraphics[width=0.3\textwidth]{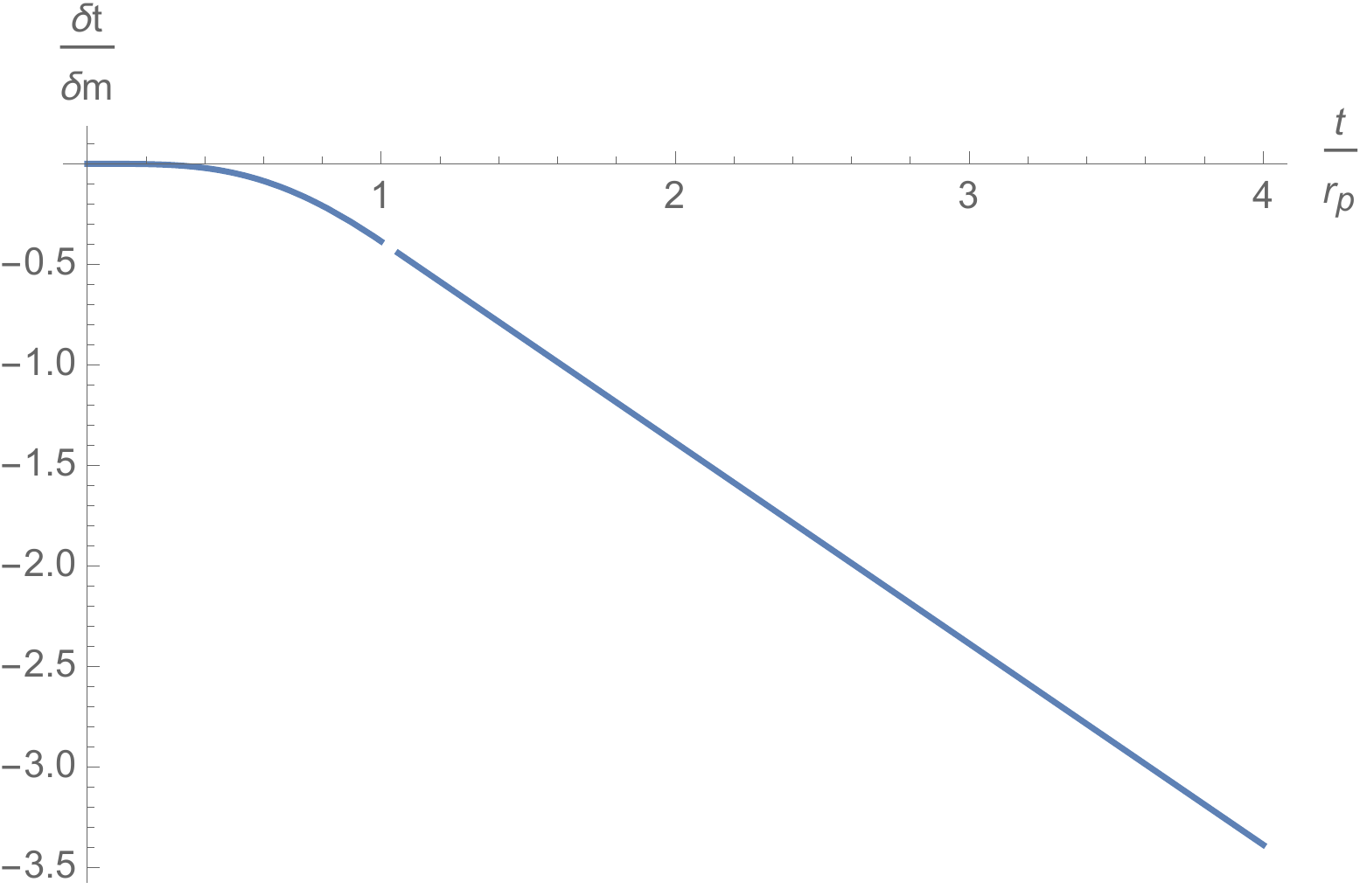}
\includegraphics[width=0.3\textwidth]{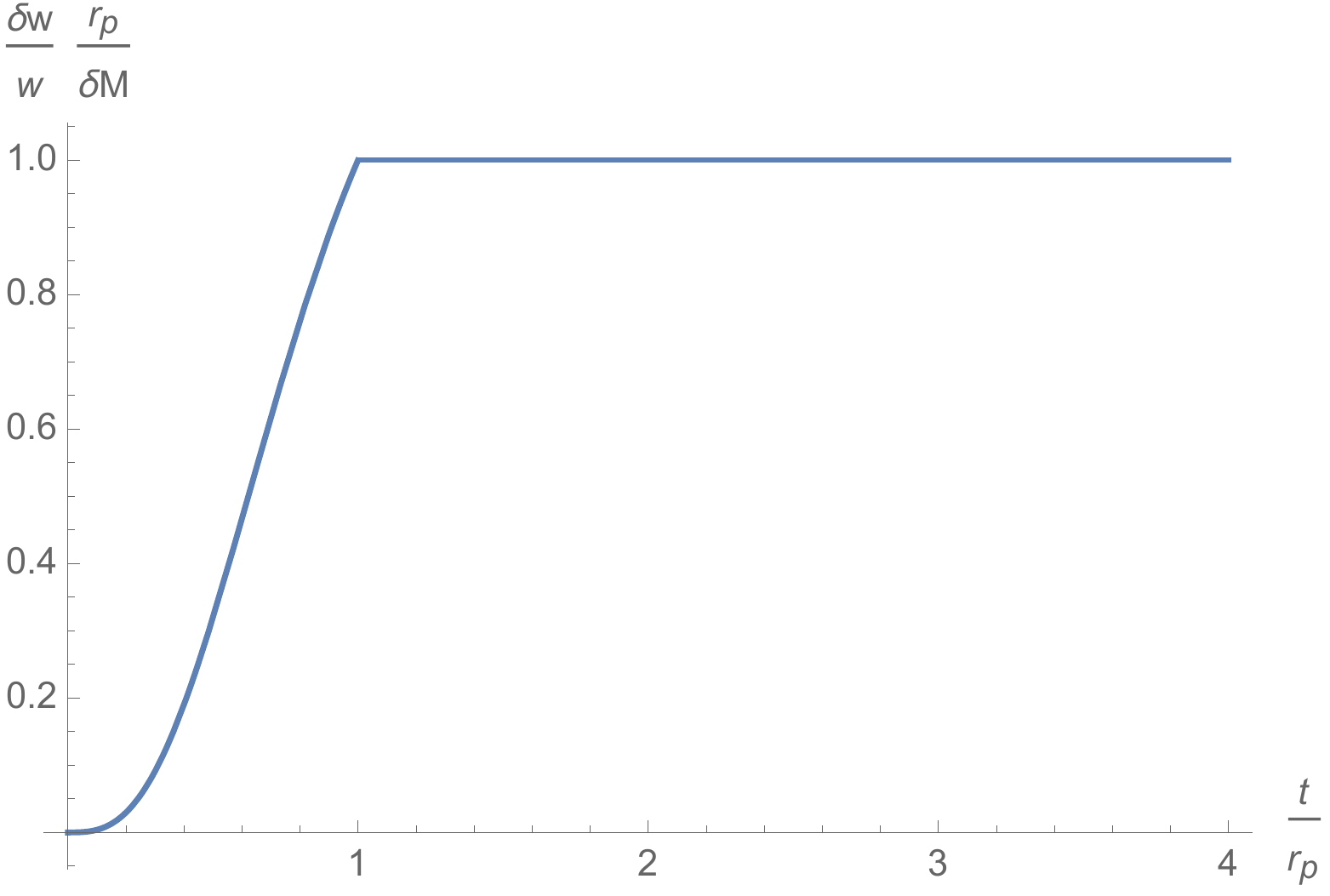}
\includegraphics[width=0.3\textwidth]{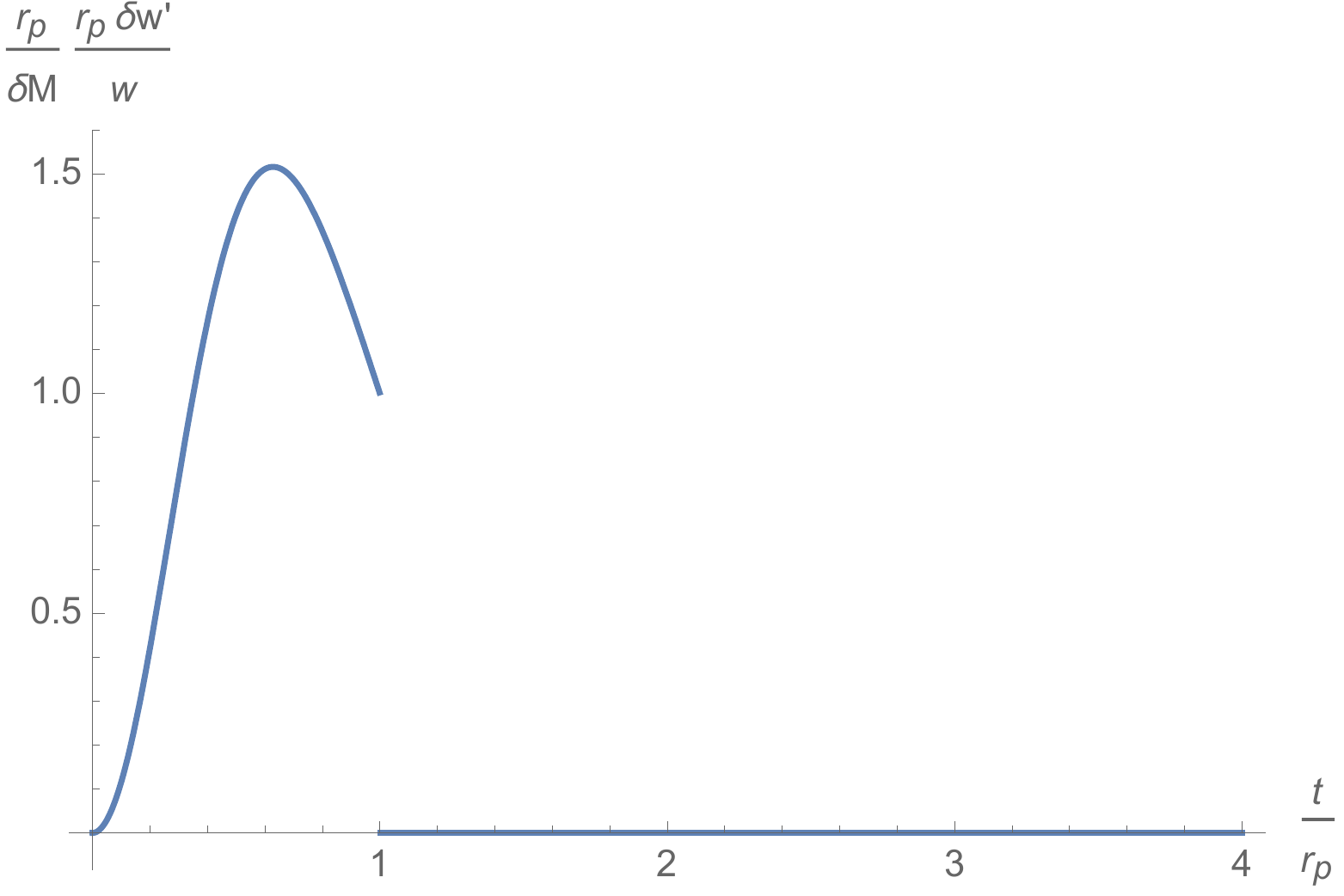} \\
\includegraphics[width=0.3\textwidth]{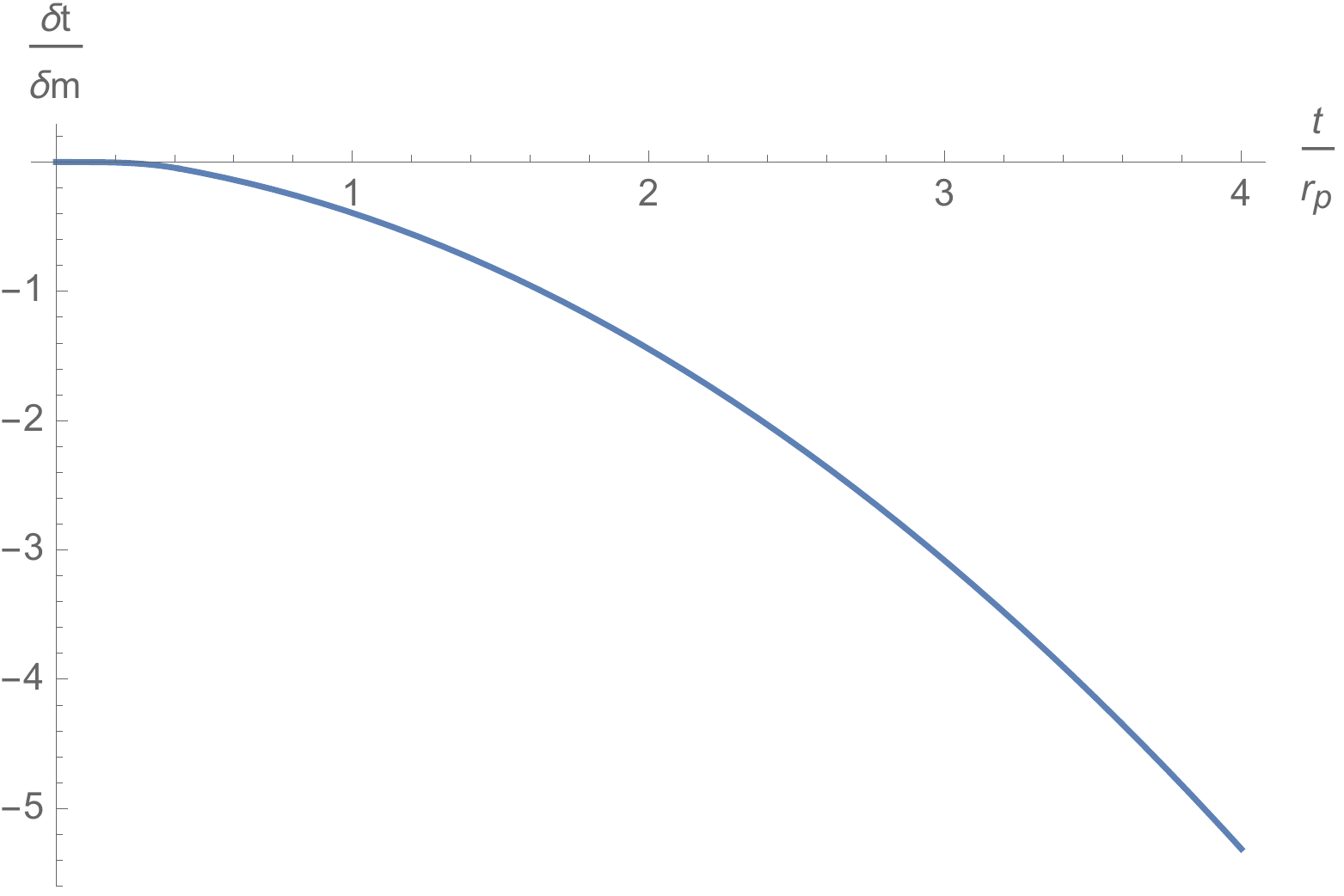}
\includegraphics[width=0.3\textwidth]{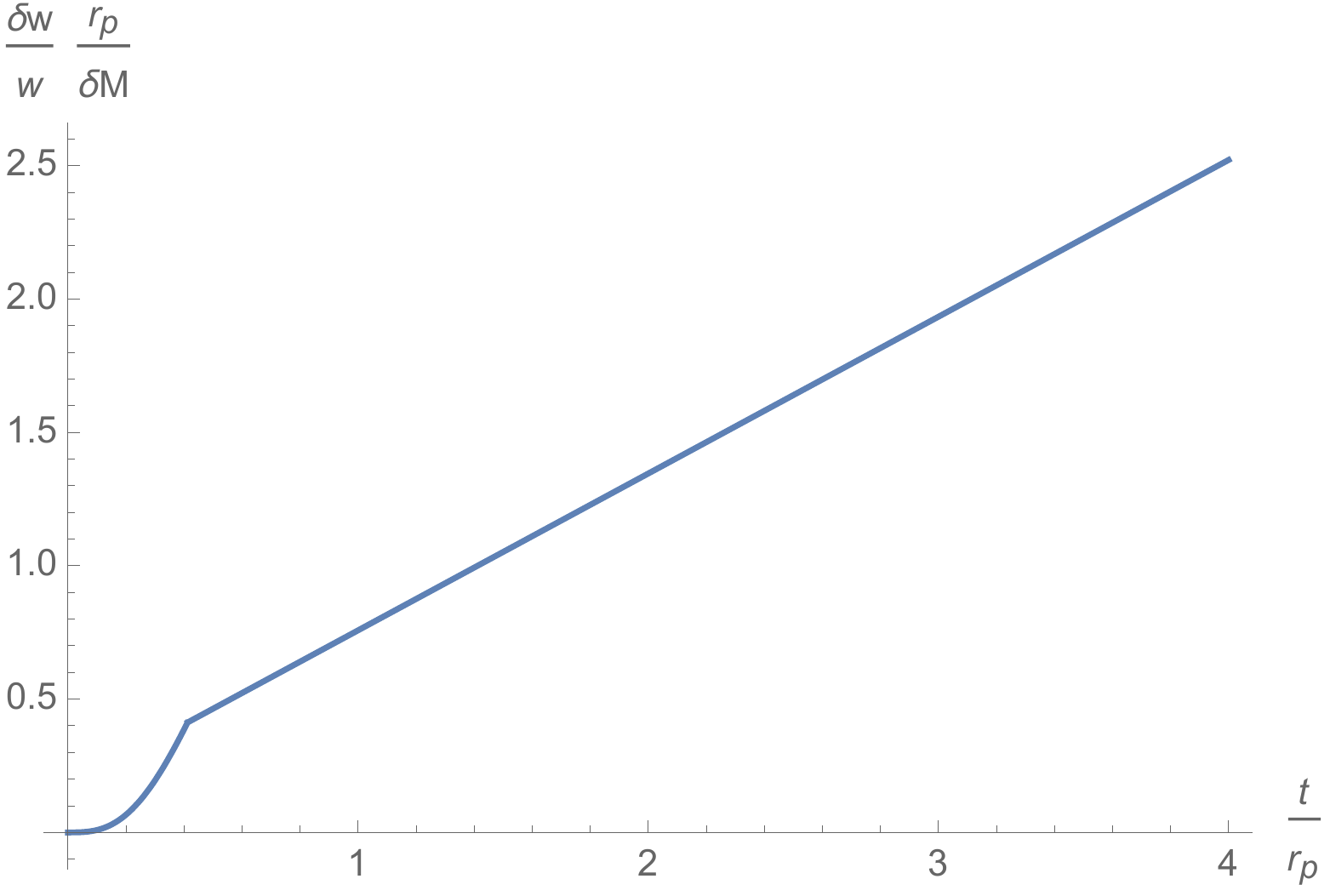}
\includegraphics[width=0.3\textwidth]{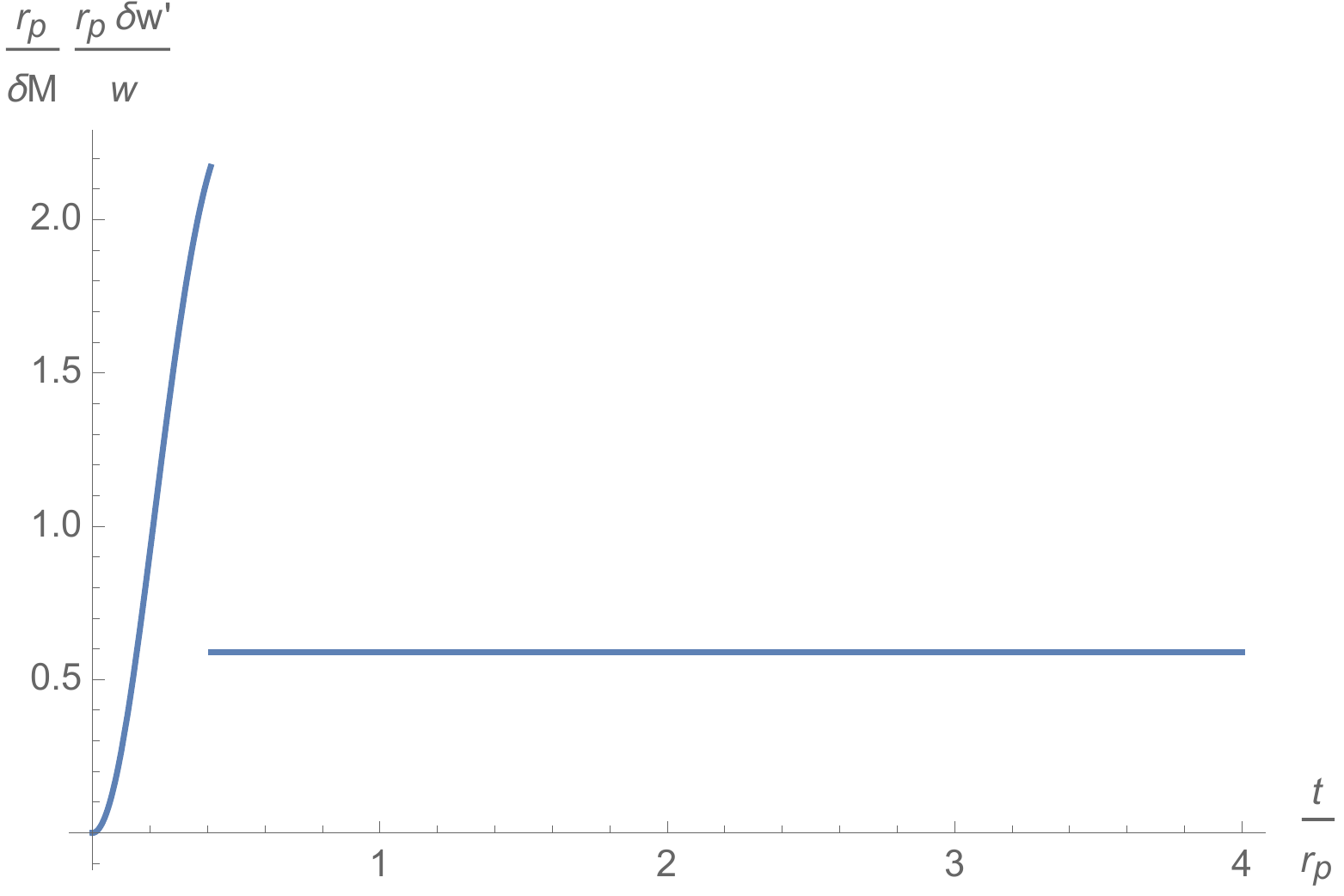}
\caption{From left to right, the pulse arrival time shift and its first and second derivatives. From top to bottom, $\theta=0.8\pi, \ 0.5\pi, \ 0.3\pi$. The distance to supernova $r_p$ and the neutrino shell mass (its effective Schwarzschild radius) $\delta M$ are used to make all quantities dimensionless.}
\label{fig-3x3}
\end{figure}

The observed value of $w_e$ cannot be exactly what we calculated, because in practice, there are many other contributions to the acceleration. For example, both the SN and the pulsar may be in the galactic center which has a relative acceleration to our spiral arm. However, if the pulsar is close enough to the SN, then we can be pretty certain that in the time scale comparable to $\delta t = r_p(1-\cos\theta)$, the SN explosion is the dominant cause of the change which starts at $t=0$. 
\begin{eqnarray}
\delta w_e(t) &=& w_0 
\frac{4\delta M t^3}{(t^2 + r_p^2\sin^2\theta)^2} ~, 
\ \ \ {\rm for} \ \ 0 < t <  r_p(1 - \cos\theta)~, \label{eq-beforeDW} \\
&=& w_0 
\left[ \frac{\delta M}{r_p}(1-\cos\theta)^2 +  t \frac{\delta M}{r_p^2}\cos\theta \right]~,
\ \ \ {\rm for} \ \ t > r_p(1-\cos\theta)~.
\end{eqnarray}
In Fig.\ref{fig-3x3}, we select three values of $\theta$ and plot the corresponding changes at the $\delta t$ timescale. Note that the most direct observable change is the shift of pulse arrival times, which can be estimated by
\begin{equation}
\delta T_{\rm arrival} \sim \delta\left(\frac{\dot{w_e}}{w_e}\right)(\delta t)^2
\sim \frac{\delta M}{r_p^2} (\delta t)^2 \sim \delta M~.
\end{equation}
The $1/r_p^2$ in gravitational force exactly cancels with the accumulation time $(\delta t)^2$. That means independent of $r_p$, a pulsar can detect this change as long as the timing accuracy is better than the light-crossing time of the effective Schwarzschild radius of the neutrino shell\footnote{In practice, we still want the pulsar to be somewhere close to the SN. Without that condition, we cannot be certain that there are no other comparable changes to acceleration during $\delta t$.}. We can estimate $\delta M\sim$ a fraction of solar mass $\sim10^{-6}$ seconds. The timing accuracy for a stable, millisecond pulsar is already better \cite{PulsarTiming}\footnote{Our conservative estimation assumes only two measurements: in the beginning and the end of the duration $\delta t$. In practice, one has $N$ measurements during the process, which increases the accuracy by another factor of $\sqrt{N}$.}.

Furthermore, the change in the first stage, Eq.~(\ref{eq-beforeDW}), as already shown in \cite{OluPie13}, depends only on the ``impact parameter'' $r_p\sin\theta$. This corresponds to the angular distance between the pulsar and SN in our sky, which can be measured on its own quite accurately. Thus fitting the arrival time shift in this stage provides a measurement of the neutrino shell mass, $\delta M$. On the other hand, there is a discontinuous change in the derivative of frequency at $t=r_p(1-\cos\theta)$, and its value depends strongly on $\cos\theta$. This can be used to calculate the difference between SN and the pulsar, which is notoriously hard to measure otherwise.

Most of the millisecond pulsars (MSP) are distributed in the galactic centre, a region with height $\sim 500$pc and radius $\sim 7.5$kpc. The mean distance from a pulsar to its nearest pulsar neighbour resulting from random catalogs of 11,000 pulsars generated using \cite{MSPpopulation} is 200.  Extrapolating to 30,000 pulsars with SKA2 \cite{SKA} and drawing SNe from the same spatial distribution (which is the most likely scenario) leads to a mean distance from a SN to the nearest timed MSP of 100 light years. It might not be unreasonable, therefore, to expect $r_p \sim30$ light years, which makes the second change in frequency observable. 


This means that as soon as we see the next SNe, we can start to follow the first stage timing shift in all MSPs. Some of them will show the predicted behavior within a few decades. By about the same time, it is likely that one or two among those will go through the second stage, having a sudden change in the second derivative of timing shifts. Given a few SNe per century \cite{SNrate06}, the detection of these effects is possible in the foreseeable future. 


\acknowledgments

We thank John Antoniadis and Ken Olum for inspiring discussions. This work is supported by the Canadian Government through the Canadian Institute for Advance Research and Industry Canada, and by Province of Ontario through the Ministry of Research and Innovation. The Dunlap Institute is funded through an endowment established by the David Dunlap family and the University of Toronto.

\appendix

\bibliography{all_active}

\end{document}